\documentclass[letterpaper,english,aps,pre,reprint]{revtex4-1}
\usepackage[T1]{fontenc}
\usepackage[latin9]{inputenc}
\usepackage{amsmath}
\usepackage{amssymb}
\usepackage{graphicx}
\usepackage{tabularx}

\makeatletter

\pdfpageheight\paperheight
\pdfpagewidth\paperwidth

\makeatother

\usepackage{babel}
\begin{document}

\title{Stickiness in generic low-dimensional Hamiltonian systems: A recurrence time statistics
approach}

\author{\v{C}rt Lozej}
\email{clozej@gmail.com}

\affiliation{CAMTP - Center for Applied Mathematics and Theoretical Physics, University
of Maribor, Mladinska 3, Maribor, Slovenia}

\date{\today}
\begin{abstract}
We analyze the structure and stickiness in the chaotic components
of generic Hamiltonian systems with divided phase space. Following
the method proposed recently in (Lozej, Robnik, Phys. Rev. E 98, 022220
(2018)), the sticky regions are identified using the statistics of
recurrence times of a single chaotic orbit into cells dividing the
phase space into a grid. We perform extensive numerical studies of
three example systems: the Chirikov standard map, the family of Robnik
billiards and the family of lemon billiards. The filling of the cells
is compared to the random model of chaotic diffusion, introduced in
(Robnik et al. J. Phys. A: Math. Gen. 30, L803 (1997)) for the description
of transport in the phase spaces of ergodic systems. The model is
based on the assumption of completely uncorrelated cell visits because
of the strongly chaotic dynamics of the orbit and the distribution
of recurrence times is exponential. In generic systems the stickiness
induces correlations in the cell visits. The distribution of recurrence
times exhibits a separation of time scales because of the dynamical trapping. 
We model the recurrence time distributions to cells inside sticky areas as a mixture of exponential distributions with different decay times.  
We introduce the variable $S$, which is the ratio between the standard deviation and the mean of
the recurrence times as a measure of stickiness. We use $S$ to globally
assess the distributions of recurrence times. We find that in the bulk
of the chaotic sea $S=1$, while $S>1$ in areas of stickiness. We
present the results in the form of animated grayscale plots of the
variable $S$ in the largest chaotic component for the three example
systems, included as supplemental material to this paper. 
\end{abstract}
\maketitle

\section{Introduction}

In generic Hamiltonian systems the phase space is divided into several
invariant components, with regular motion on some and chaotic motion
on others \citep{LichLiebBook}. These kinds of systems are usually
called mixed-type or systems with divided phase space. The exact border
between the chaotic sea(s) and the regular components is hard to determine
because of the typically infinite hierarchy of islands of stability embedded
in the chaotic sea. The chaotic sea constitutes what is known as a "fat fractal" \citep{umberger1985}.  
In systems with dimension two or lower invariant tori
strictly separate the regular and chaotic components. In higher dimensions
an invariant torus cannot strictly separate the phase space, making the analysis
even more complicated. Even in the two-dimensional case very few systems
where the border between the regular an chaotic parts can be exactly
determined are known. Examples are the mushroom billiards introduced
by Bunimovich \citep{Bun2001} and the peiecewise linear symplectic
maps  \citep{wojtkowski1981, wojtkowski1982,  MalovrhProsen2002, altmann2006}.
However, these examples have the drawback of having only a small number
of specially constructed islands of stability, in contrast to typical Hamiltonian
systems, where an infinite island-around-island structure is usually
present. Recently, a way of approximating a generic system with divided
phase was proposed \citep{BunCasProVid2019} in order to facilitate
a more rigorous analysis. The generic system is approximated by a
sequence of systems with a finite number of islands that are sub-islands
of the initial system. The approach was demonstrated for a class of
two-dimensional billiards. The intricacies of transport in generic
Hamiltonian systems remain a long standing open problem.

In systems with divided phase space, transport in the chaotic component
is strongly influenced by the various structures embedded in it. A recent review of the theory of transport 
is given in Ref. \citep{meiss2015thirty}.  
These systems commonly exhibit the phenomenon known as stickiness \citep{contopoulos1971,shirts1982}.
A good introduction to the topic of stickiness is provided by Refs. \citep{contopoulos2010a}
and \citep{bunimovich2012}. It is common for chaotic orbits to stick
to islands of stability for extended periods of time. This is due
to the presence of cantori, which are invariant cantor sets surrounding
the islands of stability that may remain after invariant curves are broken
by perturbation \citep{percival1979,mackay1984a,mackay1984b}. A chaotic
orbit with an initial condition near the last invariant curve of an island
of stability may become trapped in the region bounded by the cantorus
for an arbitrarily long time before finally exiting through one of
the holes in the cantorus into the larger chaotic sea (see Refs. \citep{contopoulos2010a,contopoulos2010b}
for a detailed description). This produces long periods of intermittent
quasi-regular motion in the chaotic orbit and results in only weakly
chaotic dynamics in the chaotic component in the sense of slow (power
law) decay of correlations \citep{zaslavsky2002} and power law tails
in recurrence time distributions. The finite time dynamics of such systems thus influences the long time transport properties.

The prevalence and universality of algebraic decays in recurrence time distributions has been a matter of intense investigation over many years \citep{karney1983,chirikov1984,hanson1985,meiss1986, zaslavsky2000, weiss2003, altmann2005, altmann2006, cristadoro2008, venegeroles2009, ceder2013, abud2013}. Most theoretical results are based on the Markov tree model \citep{hanson1985, meiss1986, cristadoro2008, venegeroles2009, ceder2013}, which predicts a power law asymptotic decay of the recurrence time distribution in generic Hamiltonian systems, taking into account the hierarchical structure of the phase space. Various numerical and theoretical studies report different values for the decay exponent ranging from 1 to 3 (a collection of the results is given in Ref. \citep{venegeroles2009}). A possible explanation is that the transition to the asymptotic regime may take an arbitrarily long time and is thus very hard to observe in numerical experiments. The existence of power law decays is also very hard to prove numerically \citep{clauset2009} as data over many orders of magnitude is needed and fluctuations often obscured by fluctuations.    
One must also stress, that a divided phase space is not a prerequisite for stickiness
and slow decay of correlations. Stickiness may also be produced by
zero measure invariant sets like families of marginally unstable periodic
orbits (MUPO) that are present also in ergodic systems (see Ref. \citep{bunimovich2012}
for a discussion). A famous example is the stadium billiard \citep{Bun1979}
where two sticky sets of MUPO are present (the so-called bouncing
ball and boundary glancing orbits) that produce power law decay of
correlations \citep{vivaldi1983}.

Stickiness may be characterized in terms of various observables. Some
examples include escape times from a given region in the phase space
\citep{contopoulos2010a,BunCasProVid2019}, finite time Lyapunov exponents
\citep{szezech2005}, recurrence plots \citep{zou2007}, recurrence
time statistics \citep{altmann2004, altmann2006, abud2013} and rotation number
\citep{santos2019}. In a recent paper \citep{lozej2018} we analyzed
the structure of the chaotic components of a single parameter family
of billiards introduced in \citep{robnik1983}. The phase space was
divided into a grid of cells and dynamics of the cell filling analyzed
in terms of the so-called random model of diffusion in chaotic components
\citep{Rob1997,ProRob1998a,RobProDob1999}. Stickiness around islands
of stability caused a slowing of the cell filling compared to the
expectation from the random model. The statistics of cell recurrence
times was studied and the standard deviation of the recurrence time
used to identify sticky areas in the chaotic component. In this paper
we use this approach to analyze the stickiness in the largest chaotic
component of several examples of generic Hamiltonian systems. The
three systems considered in this paper are the Chirikov standard map
\citep{chirikov1971}, the above mentioned family of billiards and
the family of lemon billiards \citep{HellTom1993}. 

The paper is organized as follows. In Sec. \ref{sec:RandomModel}
we present the method of analyzing the stickiness in the chaotic component
in terms of recurrence times on a grid of cells dividing the phase
space. We introduce the random model of diffusion in chaotic components
and discuss its implications on the statistics of recurrence times. we discuss the relationships between recurrence times escape times and transit times. We introduce the hyperexponential distribution to model the distributions of recurrence times in sticky areas.
We introduce the variable $S$ that is the ratio between the standard
deviations and the mean of the recurrence times as a means of identifying
stickiness. In Sec. \ref{sec:StandardMap} we apply the method to
the standard map and calculate the size of the largest chaotic component
and analyze the stickiness of the various embedded structures for
a large range of parameter values. In Sec. \ref{sec:Billiards} we
do the same for two families of billiard systems, the Robnik billiards
and the lemon billiards. In Sec. \ref{sec:Discussion} we discuss
the results and draw our conclusions. The supplemental material of
this paper contains animations of the phase spaces of the studied
systems. The animations show in terms of the variable $S$ how the
structure and stickiness of the largest chaotic component changes
as a function of the parameter. 

\section{The random model and cell recurrence times\label{sec:RandomModel}}

In this paper we use the moments of the distribution of recurrence
times to identify sticky areas in the chaotic component, following
the approach proposed in our recent paper \citep{lozej2018}. The
approach is based on the \emph{random model of diffusion} in chaotic
components, first introduced in Ref. \citep{Rob1997} as a model of
transport in ergodic chaotic systems and extended in Refs. \citep{ProRob1998a,RobProDob1999}
for systems with divided phase space and systems with several weakly
coupled ergodic sub-components. The main idea of our approach is to
use a single chaotic orbit to generate the recurrence time data. The benefit is that no prior knowledge of the structure of the phase space is needed, only a single initial condition for the chaotic orbit.
The idea of mapping the chaotic component by using a single chaotic orbit was used already by Umberger and Farmer \citep{umberger1985}. The procedure is as follows. Let $M$ be the phase space (surface of section), $f:M \rightarrow M$ the mapping and $\mu$ the invariant measure of our discrete dynamical system and $\mu(M)=1$. We divide $M$ into a grid of $L\times L$ rectangular cells. We select
a single initial condition in the chaotic component and iterate the
orbit $T$ times. At each iteration the orbit visits one of the cells. We will
refer to cells that are visited by the orbit at least once as \emph{filled
cells} and those that are never visited as \emph{empty cells. }Eventually
the chaotic orbit will explore all of the available phase space and
the filled cells will cover the chaotic component $C$. The empty cells
belong to other invariant components. 

The random model assumes that in strongly chaotic systems the cell
visits are completely uncorrelated, independent from previous cell
visits, they constitute a Poisson process. With this assumption the proportion of filled cells $\chi$
at time $T$ (number of iterations) follows the exponential law
\begin{equation}
\chi\left(T\right)=\chi_{c}\left(1-\exp\left(-\frac{T}{N_{c}}\right)\right),\label{eq:CellFill}
\end{equation}
where $\chi_{c}=\mu(C)$ is the measure of the chaotic component and $N_{c}=\chi_{c}L^{2}$
is the number of cells available to the chaotic orbit. This exponential
law was first derived only for ergodic systems where $\chi_{c}=1$
and has been shown to excellently describe real data (see Ref. \citep{Rob1997}
for details). In systems with divided phase space the basic assumption
of uncorrelated cell visits does not hold in general. If the system
exhibits stickiness the cell visits within the sticky areas become
correlated. As we shall see in the numerical examples the cell filling
is slowed by stickiness. If there are no sticky areas the cell filling
is well described by Eq. \eqref{eq:CellFill} even in systems with
divided phase space.

To quantify the effects of stickiness we measure the recurrence times
in each of the cells visited by the chaotic orbit. Let $A \subset M$ for instance one of the cells. The first recurrence
time to $A$ for a point $a \in A$ is defined as the number of iterations an orbit needs
to return to the same cell for the first time,
\begin{equation}
\tau_A=\underset{t>0}{\mathrm{min}}\{t:f^t(a) \in A\}. \label{eq:RecTime}
\end{equation}
We are interested in the probability distributions of recurrence
times 
\begin{equation}
P\left(j\right)=\frac{\mu(\{a \in A:\tau_A(a)  = j\})}{\mu(A)} \label{eq:RecTimeDist}
\end{equation}
and its moments to the cells dividing the phase space.
The idea of using the mean recurrence time to probe the size of the accessible area was first given by Meiss in Ref. \citep{meiss1997} together with several results concerning the relationships between the transit, exit and recurrence times.
In ergodic components the mean recurrence time
is given by the Kac lemma \citep{Kac1959} which states
\begin{equation}
\langle \tau_A \rangle_A = \frac{\mu(A_{acc})}{\mu(A)}, \label{eq:KacLemma}
\end{equation}
where the angled brackets denote the phase space average over $A$ and $A_{acc} \subset M$ is the subset of the phase space accessible to orbits starting from $A$.
Taking one of the cells in the chaotic component as $A$, its accessible set is the chaotic component and the mean recurrence time to the cell is equal to $N_{c} = \chi_{c}L^{2}$. 
In numerical experiments the measured mean recurrence times are distributed
normally around the theoretical mean because of finite sample size
effects (see Ref. \citep{lozej2018} for details). Additionally, if
we assume the premise of the random model that the cell visits are
completely uncorrelated (a Poisson process), the probability that a cell is visited after
any number of iterations is equal, resulting in an exponential probability density function
of recurrence times
\begin{equation}
P\left(\tau\right)=\frac{1}{N_c}\exp\left(-\frac{\tau}{N_c}\right),\label{eq:ExponentialDist}
\end{equation}
to each individual cell. This is the result we expect in systems with strong mixing properties \citep{hirata1999}. A common way of describing the recurrence time distributions is also in terms of the survival function or complementary cumulative distribution i. e. the probability that the recurrence time is greater than $t$, $W(t) = \sum_{\tau >t} P(\tau)$ referred to as the Poincar\'{e} recurrence time distribution by some authors \citep{chirikov1984,chirikov1999}. In the Poissonian case
\begin{equation}
W\left(t \right)=\exp\left(-\frac{t}{N_c}\right).\label{eq:ExponentialDist}
\end{equation}
However, the assumption of uncorrelated cell visits does not hold for areas of stickiness. If the cell is located inside an area of stickiness, the orbit is likely to visit the same cell again before escaping into the grater chaotic sea because of the dynamical trapping. As a consequence the recurrences when the orbit leaves the sticky area may happen on a vastly different time scale then when it stays inside for the entire period of recurrence. The strength of the trapping inside the sticky area may be quantified by the escape time. Let us again consider a subset $A$ of the phase space. The \emph{exit set} of $A$ is the set of all points that exit $A$ after one iteration $E = A \setminus f^{-1}(A)$. Similarly the \emph{entry set} is the set of all points that enter $A$ in one iteration $I = A \setminus f(A)$. The union of the exit and entry sets is called a turnstile. The escape time is the time needed for an orbit starting in $a \in A$ to leave
\begin{equation}
t^{esc}_A=\underset{t>0}{\mathrm{min}}\{t:f^t(a) \in M \setminus A\}. \label{eq:EscTime}
\end{equation}
Analogously the entry time is the time needed to enter the set from outside
\begin{equation}
t^{ent}_A=\underset{t>0}{\mathrm{min}}\{t:f^{-t}(a) \in M \setminus A\}. \label{eq:EntTime}
\end{equation}
Orbits starting in the exit set $a \in E$ will escape after one iteration $t^{esc}_E=1$ and $t^{ent}_I=1$. Similarly orbits starting within $A$ but not in $E$ will recur to $A$ after one iteration meaning the recurrence time is $\tau_{A \setminus E} = 1$. The time to transit an area is $t^{trans}_A = t^{ent}_A + t^{esc}_A -1$. The recurrence time to $A$ is essentially the transit time of the accessible area $A_{acc}$.
The above relations as well as many other useful results are derived in Refs. \citep{meiss1997, meiss2015thirty}.  Let us now consider the recurrence times to a cell embedded inside a sticky area. The orbit starts from inside the cell and must first escape through the exit set of the cell. Then the orbit is trapped inside the sticky area for the duration of the typical escape time. The orbit may visit the cell before escaping. The rate at which this happens on average depends on the measure of the sticky area and the measure of the entry set of the cell. The orbit may also escape the sticky region before returning to the cell. The escape times from sticky areas are again related to the measures of the sticky area and its exit set. The flux trough the bordering cantori may in some cases be estimated analytically using transport theory \citep{mackay1984a,mackay1984a,meiss2015thirty}. The orbit then spends a typical transition time in the larger chaotic sea. Depending on the structure of the phase space the orbit may also visit other sticky areas each with its own typical entry and escape times. After transitioning back to the original sticky area the orbit may again return to the cell. The typical time scales for the different possible transitions may vary greatly. We may thus expect to see several typical time scales in the recurrence time distributions to the cells inside the sticky regions.    

Probing the distribution of recurrence times can only be feasibly
done in a few selected cells. Our previous numerical results \citep{lozej2018}
as well as those presented in this paper show that the distributions of
recurrence times do indeed follow the exponential law \eqref{eq:ExponentialDist}
in the bulk of the chaotic sea, far away from any islands. On the
other hand the distributions in cells located in the sticky areas exhibit
a short time peak followed by an exponential tail. Similar results have been found
by Altmann et. al. \citep{altmann2004} in one dimensional chaotic
maps. We propose to model the separation of time scales introduced by the trapping inside sticky regions with a mixture of exponential distributions with $n$ different time scales (known also as the hyperexponential distribution). We model the survival function as 

\begin{equation}
W\left(t \right)=\sum_{i=1}^n{p_i\exp\left(-\frac{\lambda_i t}{N_c}\right)},\label{eq:HyperExponentialDist}
\end{equation}
where $p_i$ are the mixture coefficients and $\sum_{i=1}^n{p_i}=1$ and $\lambda_i$ are dimensionless parameters characterizing the relevant time scales in terms of the mean recurrence time $N_c$. The mean $m$ and variance $\sigma^2$ of the distribution \eqref{eq:HyperExponentialDist} are given by 

\begin{align}
m &= N_c \sum_{i=1}^n{\frac{p_i}{\lambda_i}}, \label{eq:mean} \\
\sigma^2 & = {N_c}^2 \sum_{i=1}^n{\frac{2}{{\lambda_i}^2}p_i} - m^2.\label{eq:variance}
\end{align}

For a global understanding of the distributions of recurrence times
their moments can be calculated for all cells iteratively with each
orbit visit. In cells where the distribution of recurrence times is
exponential \eqref{eq:ExponentialDist}, the standard deviation is
equal to its mean $\sigma=N_{c}$. Due to the Kac lemma the mean 
recurrence time is $m = N_c$ also in sticky cells, giving the relation $\sum_{i=1}^n{\frac{p_i}{\lambda_i}}=1$ for the parameters of distribution \eqref{eq:HyperExponentialDist}. If the number of exponential components is $n>1$ the standard deviation is increased $\sigma>N_c$ \citep{papadopolous1993queueing}.  
The variable $S=\sigma/N_{c}$ (the coefficient of variation) is thus very useful for identifying sticky cells. In areas of the chaotic
component with strong chaos and no stickiness $S=1$, indicating an
exponential distribution of recurrence times, while $S>1$ indicates
stickiness. The larger the value of $S$ the stronger the stickiness.
However, a comment must be made in regard to the convergence of $S$
at infinite times. In our model we only take into account exponentially decaying distributions. Other types of distributions may also be applicable. Of particular interest are recurrence time distributions with algebraic decays.
The variable $S$ would diverge if the distribution of recurrence times exhibits sufficiently strong power law tails. If asymptotically the distribution decays as $P\left(\tau\right)\sim \tau ^{-\gamma} $,
the second moment diverges if $\gamma < 3$. In the numerical case
this would mean that the value of $S$ would keep increasing with
the number of orbit iterations. This may limit the applicability of
the variable $S$ in the infinite time limit, but would still provide
valuable information about the stickiness for finite times. We note that when comparing results for the exponents with other papers one must take care as some authors use $P(\tau)$ and others $W(t)$ as the observed distribution.

\section{Results for the standard map\label{sec:StandardMap}}

\begin{figure}[h]
\begin{centering}
\includegraphics{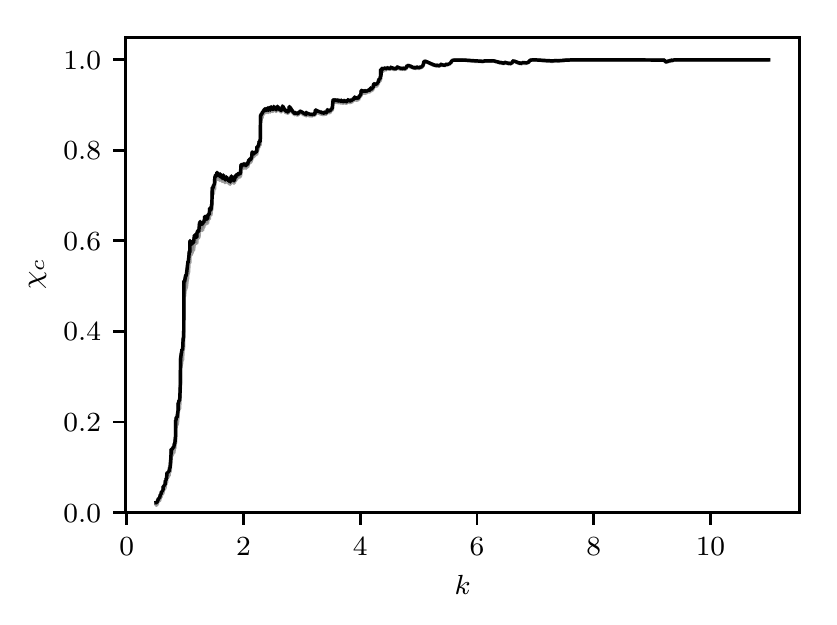}
\par\end{centering}
\caption{\label{fig:StandardMapArea}The relative size of the largest chaotic
component $\chi_{c}$ in the standard map as a function of $k$. The
grid size is $L=1000$ and the orbit was iterated $T=10^{10}$ times.
The gray area shows the error estimated from the number of cells bordering
regular components and is hardly visible. See the animation in the
supplemental material for the corresponding $S$-plots.}
\end{figure}

The Chirikov standard map \citep{chirikov1971} is one of the most
well studied 2D area-preserving mappings and is applicable to many
areas of physics (for a review see Ref. \citep{Chirikov:2008}). The
mapping is given by
\begin{gather}
p^{\prime}=p+k\sin(x),\\
x^{\prime}=x+p^{\prime},\nonumber 
\end{gather}
where we consider the variables on a torus $\left(x,\,p\right)\in\left[0,\,2\pi\right]\times\left[0,\,2\pi\right]$,
taking both variables $\mathrm{mod}\,2\pi$ and the prime symbol denoting
the variables after one map iteration. The parameter $k$ controls
the degree of chaos in the system. In Fig. \ref{fig:StandardMapArea}
we show how the overall size $\chi_{c}$ of the largest chaotic component
changes with the parameter value. This is calculated by counting the
number of filled cells after $T=10^{10}$ mappings.  The chaotic component is a "fat fractal" \citep{umberger1985} making an accurate estimate for its area difficult. In the infinite time limit the filled cells cover the chaotic sea. Naturally, the cells on the border of the chaotic sea must partially cover also the other invariant components. Taking all of the area of the filled cells as belonging to the chaotic component thus overestimates its true area. In the worst case the border cells barely touch the chaotic component and in reality barely contribute to the real area of the chaotic sea and thus counting them as filled overestimates the true area of the chaotic component. In the numerical data we define a border cell as one that has at least one empty neighbour. By this definition all features that are smaller than one cell, like for instance tiny islands of stability, are missed. The maximum error for the (lower bound of the area) is estimated by taking the border cells as empty. The area changes relatively smoothly with the parameter value but not
monotonically, with many oscillations when the various islands
are destroyed. 

\begin{figure}[h]
\begin{centering}
\includegraphics[width=1\columnwidth]{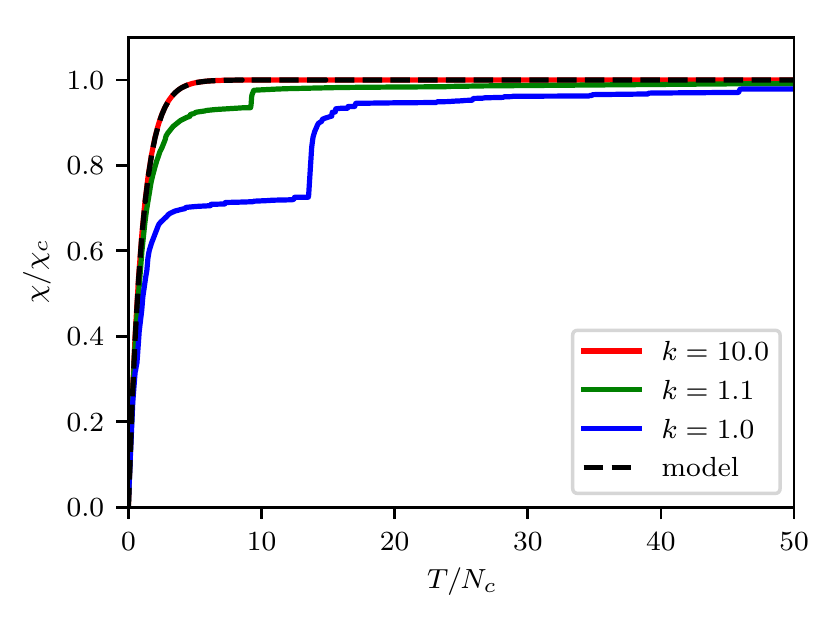}
\par\end{centering}
\caption{\label{fig:CellFillStandardMap}The proportion of filled cells (normalized
by the proportion of chaotic cells) with the number of standard map
iterations (normalized by the number of chaotic cells). The colored
lines show the cell filling curves for different values of the parameter
$k$. Each curve shows the cell filling for a a single chaotic orbit.
The dashed black curve shows the random model prediction. $L = 1000$.}

\end{figure}

Below the critical value $k<k_{c}$ invariant curves limit transport
in the $p$ direction and the phase space features several separate
chaotic components of significant size. At the critical value $k_{c}\approx0.9716$
\citep{greene1979,mackay1983renormalization} the chaotic components
merge into a single one where the variable $p$ can take all values. At this value the so called golden invariant circles are broken.
However, for $k\gtrsim k_{c}$ cantori, that remain after the last
invariant curve is destroyed, may severely impede transport from one
part of the chaotic component to the other. This may very clearly
be seen in the cell filling curves $\chi(T)$ using the procedure
described in Sec. \ref{sec:RandomModel}. In Fig. \ref{fig:CellFillStandardMap}
we show the cell filling for three values of $k$. The cell filling
closely follows the random model prediction at $k=10$ where the phase
space is practically entirely filled by one chaotic component and
no islands are visible. In the other cases the cell filling is
slowed and distinct step-like features appear. The steps signify that
the orbit is confined in some area of phase space for some time before
eventually finding its way through one of the holes in the cantorus.
The increase in the number of visited cells is therefore halted for
the duration of the trapping. The size of the steps gives an indication
of the relative sizes of the areas of phase space separated by the
cantori. Similar large steps in the cell filling at $k \gtrsim k_c$ were found also by Meiss in Ref. \citep{meiss1994} because of the trapping of the orbits by the golden cantorus.  

\begin{figure}[h]
\begin{centering}
\includegraphics[width=1\columnwidth]{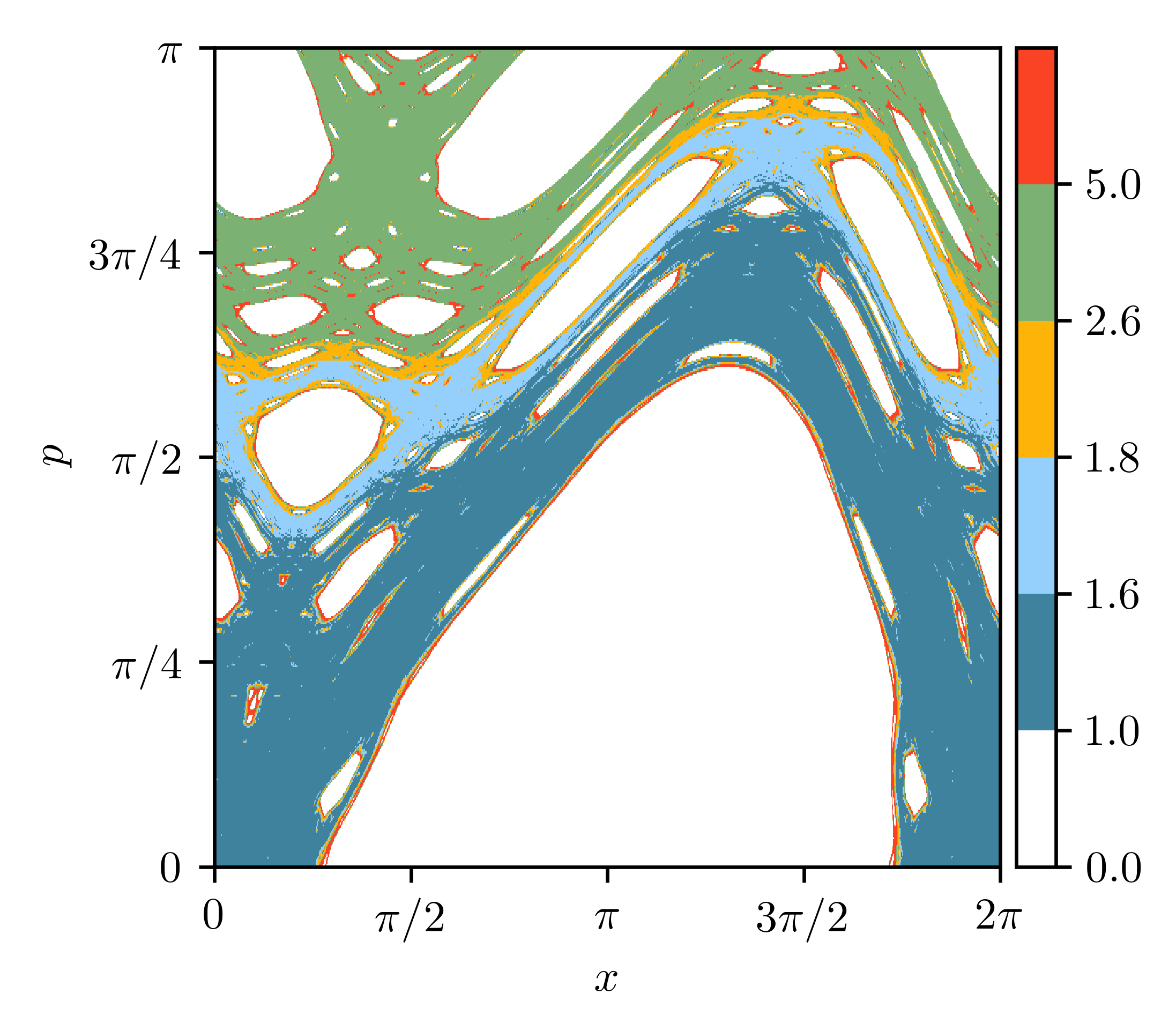}
\par\end{centering}
\caption{\label{fig:StandardMapSubcomponents}The decomposition of the phase space of the standard map at $k=1$ into sub-components using the variable $S$. Due to symmetry only half of the phase space is shown. The color changes at selected values of $S$ are indicated on the color bar. Components not belonging to the largest chaotic sea are shown in white. Cantori impede transport between the different sub-components producing different values of $S$ in each. $S$ is roughly uniform in each sub-component. The largest values of $S$ (red) indicating the strongest stickiness may be seen around many of the islands of stability.}
\end{figure}

The areas of stickiness may be found by examining the cell recurrence
times in the chaotic component. In Fig. \ref{fig:StandardMapSubcomponents}
we show a color plot of $S=\sigma/N_{c}$ on a grid of $1000\times1000$
cells for the largest chaotic component of
the standard mapping at $k=1.0$, right above the critical value. 
The values of $\sigma$ and $N_{c}$ are determined numerically
at $T=10^{10}$ mappings for each visited cell. Similar results are
obtained if the value for the mean is calculated from the numerically
determined $\chi_{c}$ using the formula $N_{c}=\chi_{c}L^{2}$ instead.
The white areas belong to other invariant components, mostly islands
of stability. Several large distinct areas of uniform values of $S$ are
visible. They stratify the phase space in the $p$ direction. In the largest one (dark blue) $S=1.5$, followed by $S=1.6$ (light blue), a thin layer with $S=2.4$ and $S=3.8$ in the second largest (green). Areas of stickiness may also be seen around several of the islands of stability with $S>5$ (red) usually increasing in several stages. Each change in color signifies a strong barrier (cantorus with small holes) that is present between the two adjacent areas. By examining the phase portrait for $k$ right below the critical
value $k_{c}\approx0.9716$ and comparing it to the $S$ plot at $k=1.0$
we see that the border between the green and yellow areas is very near the last
spanning invariant curve (the golden invariant circle) before its destruction (not shown). The cantorus left
in its place at $k=1.0$ causes the trapping of the orbit in one area
or the other. The value of $S$ may be used to quantify the relative strength of the stickiness when comparing the different sticky areas.

\begin{figure*}[]
\begin{centering}
\includegraphics[width=1\textwidth]{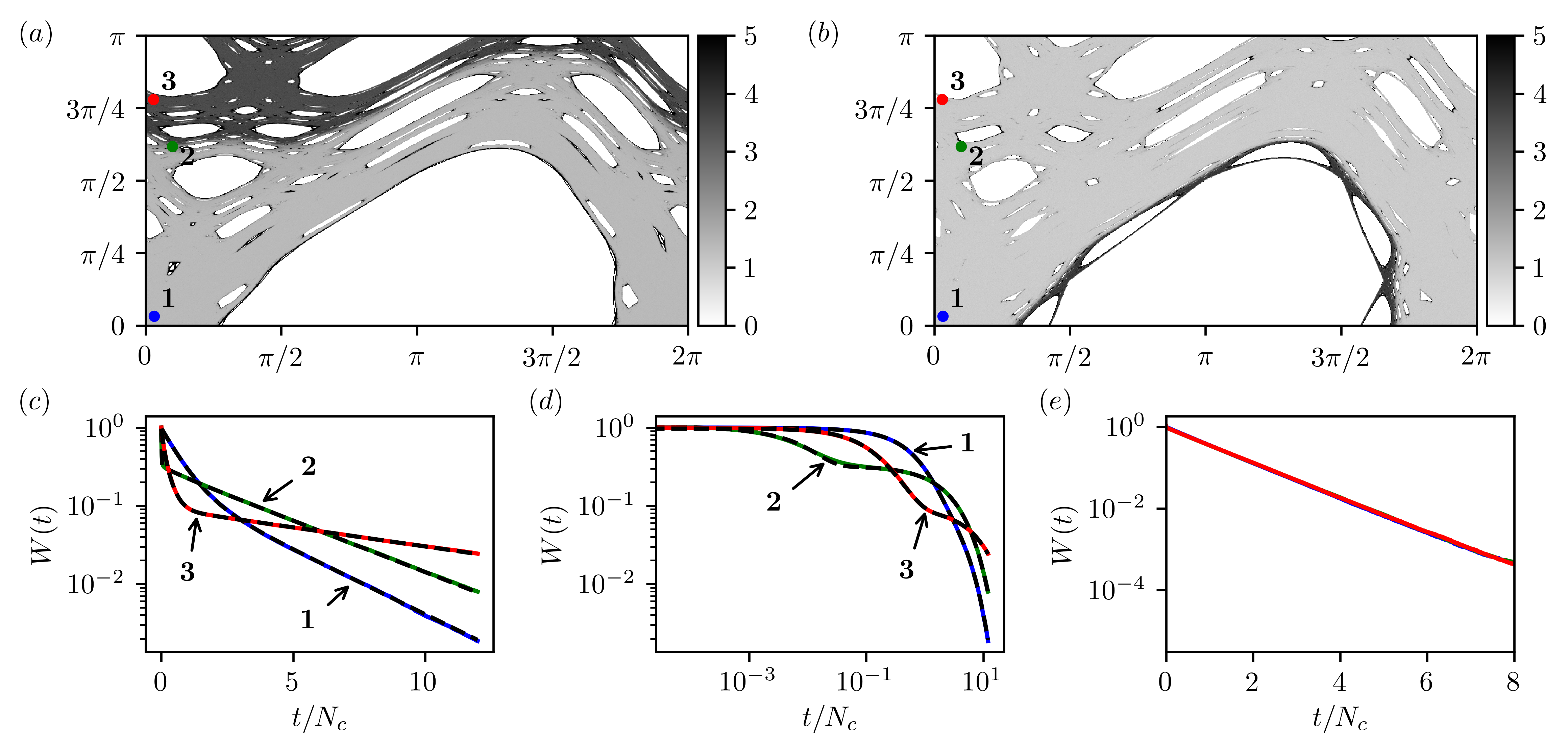}\caption{\label{fig:StandardMapPortraits}The $S$-plots of the largest chaotic
component in the standard map at $T=10^{10}$, $L=1000$ for (a) $k=1$ and (b) $k=1.1$. Due
to symmetry only half of the phase space is shown $\left(x,\,p\right)\in\left[0,\,2\pi\right]\times\left[0,\,\pi\right]$. The color bar shows the corresponding value of $S$.
Darker areas indicate stickiness. White areas belong to separate invariant
components. The survival functions of recurrence times for three cells
are shown the lower panels for $k=1$ in the log-lin plot (c) and log-log plot (d) and for $k=1.1$ in the log-lin plot (e).
The recurrence time is given in units of the mean. The black dashed curves show the fitted hyperexponential distributions \eqref{eq:HyperExponentialDist}. The positions of
the cells are shown as a dot of the corresponding color (online version)
and number. The cell coordinates are 1: $(0.1,0.1)$, 2: $(0.31,1.94)$, 3: $(0.09,0.2.45)$.  The distributions in the $k=1.1$ case are all exponential
and overlap with a decay rate of $1/N_c$. The distribution data is generated from more than $10^5$ recurrences to each cell at $T=10^{11}$, $L=1000$.}
\par\end{centering}
\end{figure*}

In Fig. \ref{fig:StandardMapPortraits}
we show gray-scale plots of $S=\sigma/N_{c}$ on a grid of $1000\times1000$
cells. In the following we shall refer to this type of plot as an
$S$-plot. In panel (a) we show again the plot for the largest chaotic component of
the standard mapping for $k=1.0$. On panel (b) the S-plot at $k=1.1$ is shown. Comparing the two we see that the phase space at $k=1.1$ is no longer stratified in the $p$ direction. The holes in the cantori have grown large enough as to no longer impede the transport and most of the chaotic component is covered by cells with
$S=1$ indicating a nearly perfect exponential distribution of recurrence
times. The largest sticky area, with $S=5$ in this case is formed
around the large island around the stable stationary point $x=\pi$,
$p=0$. This is responsible for the small step in the cell filling
for $k=1.1$ as seen in Fig. \ref{fig:CellFillStandardMap}. Thin
sticky areas may also be seen around many of the small islands. 

The survival functions of recurrence times in three selected cells are
presented in the lower panels of Fig. \ref{fig:StandardMapPortraits}.
The positions of the cells in the phase space are shown as labeled
(colored in online version) dots in panels (a) and (b). In panel (c)
we show the survival function at $k=1$ in the three distinct areas with
different values of $S$ described above. The first cell (blue) is
located in the large sub-component with $S=1.5$. The second in the thin layer with $S=2.4$. The third cell (red) is located in
the upper sub-component $S=3.8$. The three survival functions are shown also in the log-log plot in panel (d).  The survival functions can effectively be modeled using the hyperexponential distribution \eqref{eq:HyperExponentialDist}. The fitting procedure is done using a version of Prony's method \citep{feldmann1997} and goes as follows. At large enough values of $t$ only the exponential with the slowest decay rate significantly contributes to the survival function. We select a cutoff point at some large enough value $t_1$ and consider only the tail of the survival function. We fit the tail with a single exponential function $f_1(t)=p_1\exp\left(-\frac{\lambda_1 t}{N_c}\right)$. We then subtract $f_1$ from the data to eliminate this contribution. We repeat the process until the desired number of time scales is reached. We used up to four exponential functions to fit the data. The parameters are given in table \ref{table:1}. Using the distribution parameters in formulas \eqref{eq:mean} and \eqref{eq:variance} we obtain results consistent with the values of $S$ taken from the S-plots. In table \ref{table:1} and the the log-log plots on panel (d) we may clearly see that the time scales of recurrences are significantly separated in the sticky cells. The short time recurrences must be associated with orbits that do not leave the sticky area before revisiting the cell. The other time scales are probably related with the orbit sticking to other substructures and transitions between them through the chaotic sea. To be able to determine the parameters analytically a thorough understanding of the structure of the sticky sets and the associated turnstiles would be needed and is in general very difficult. At $k=1.1$ the survival functions in all three cells are exponential (panel (e)) with decay rate $1/N_c$.

\begin{table}[h]
\begin{center}
\begin{tabularx}{0.8\columnwidth} { | >{\centering\arraybackslash}X | >{\centering\arraybackslash}X | >{\centering\arraybackslash}X | >{\centering\arraybackslash}X | >{\centering\arraybackslash}X | }
 \hline
 \multicolumn{5}{| c|}{Cell 1} \\
 \hline
$\lambda_i$ &0.380 & 1.55 & 53.0 & n/a \\
$p_i$ &0.18 & 0.80 & 0.02 & 0 \\
\hline
 \multicolumn{5}{|c|}{Cell 2} \\
 \hline
$\lambda_i$ & 0.326 & 2.01 & 105 & 1210 \\
$p_i$ &0.29& 0.04 & 0.59 & 0.10 \\
\hline
 \multicolumn{5}{|c|}{Cell 3} \\
 \hline
$\lambda_i$ &0.112 & 3.69 &10.2 & 142 \\
$p_i$ &0.09 & 0.46 & 0.40 & 0.036 \\
\hline

\end{tabularx}
\caption{Table of parameter values for the hyperexponetial distributions fitted to the survival functions for $L=1000$ presented in Fig. \ref{fig:StandardMapPortraits} (c) and (d) and Fig. \ref{fig:Svsgrid} (b).}
\label{table:1}
\end{center}
\end{table}

\begin{figure}[]
\begin{centering}
\includegraphics[width=1\columnwidth]{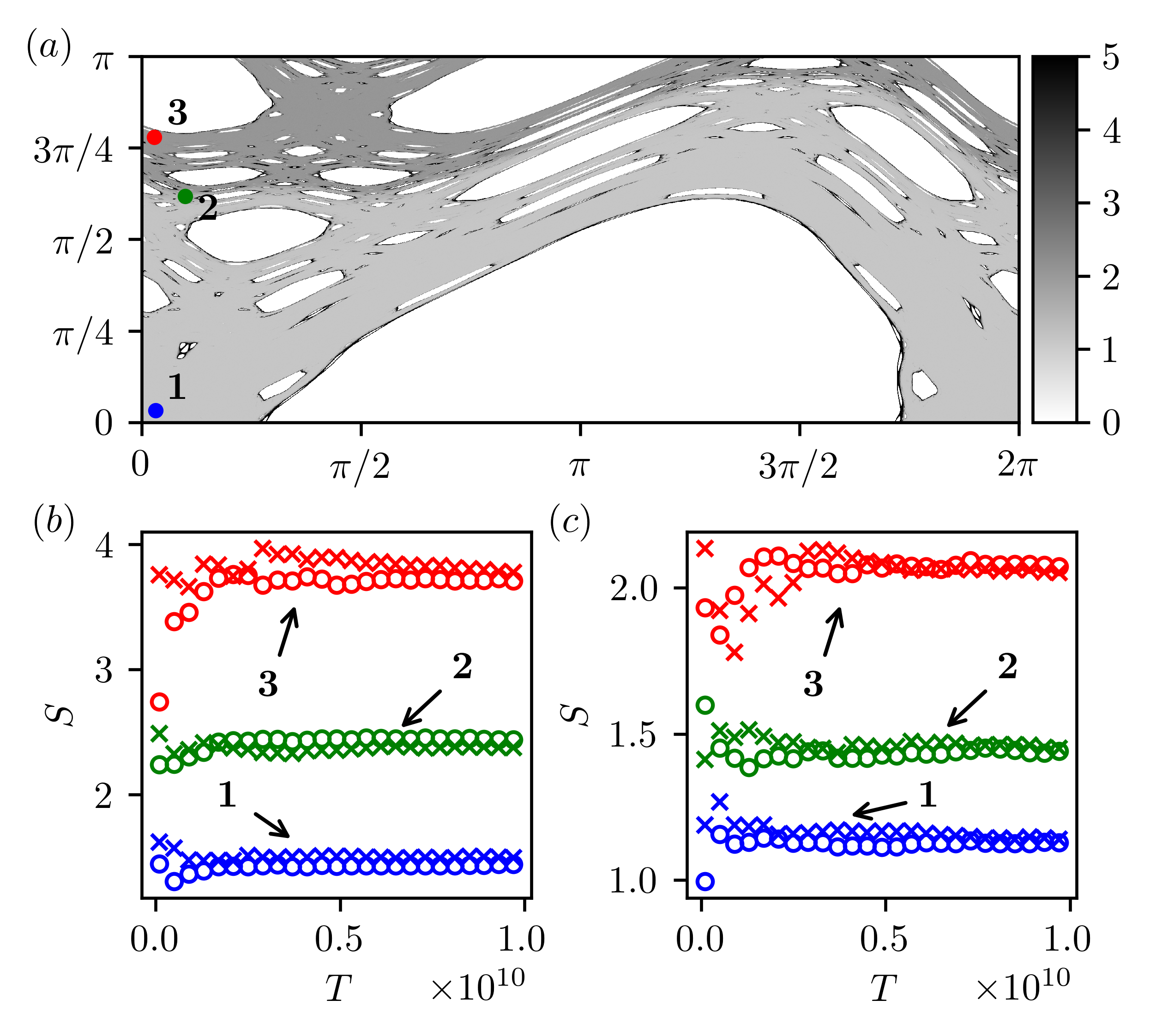}\caption{\label{fig:Svstime} (a) The $S$-plot of the largest chaotic
component in the standard map at $k=1$ at $L=2000$ and $T=10^{10}$. Panels (b) and (c)  show $S$ as a function of $T$ in three cells. The positions of the cells are indicated in panel (a) and are the same as in Fig. \ref{fig:StandardMapPortraits}. (b) $L=1000$, (c) $L=2000$, $T=10^{10}$. The circles show the results for the orbit starting at $(0.1,0.1)$ and the crosses at $(0.11,0.11)$. The same two orbits are shown in both panels.} 
\par\end{centering}
\end{figure}

The numerical stability of the value of $S$ in the three cells is analyzed in Fig. \ref{fig:Svstime}. In panel (a) we show the $S$-plot at $k=1.0$, $L=2000$ and $T=10^{10}$. Comparing this with the plot at $L=1000$ in Fig. \ref{fig:StandardMapPortraits} (a) we see the results are qualitatively the same but the values of $S$ are lower in sticky cells than they were at $L=1000$. Cells with greater stickiness still exhibit higher values of $S$. In non-sticky cells $S=1$ regardless of the grid size. In panels (b) and (c) we plot the value of $S$ as a function of the number of iterations for two orbits one with the initial condition $(x,p) = (0,1,0.1)$ and the other $(0.11,0.11)$. We compare the results for the two orbits at different grid sizes. Panel (b) shows the results at $L=1000$ and panel (c) at $L=2000$. After a transient regime the value of $S$ stabilizes and both orbits give similar results. The transient is more pronounced in cell 3 and the fluctuations are larger as the orbits need to cross the strong barriers of the golden cantori to reach this region. The distributions of recurrence times are therefore quite stable after the transient regime.  

\begin{figure}[h]
\begin{centering}
\includegraphics[width=1\columnwidth]{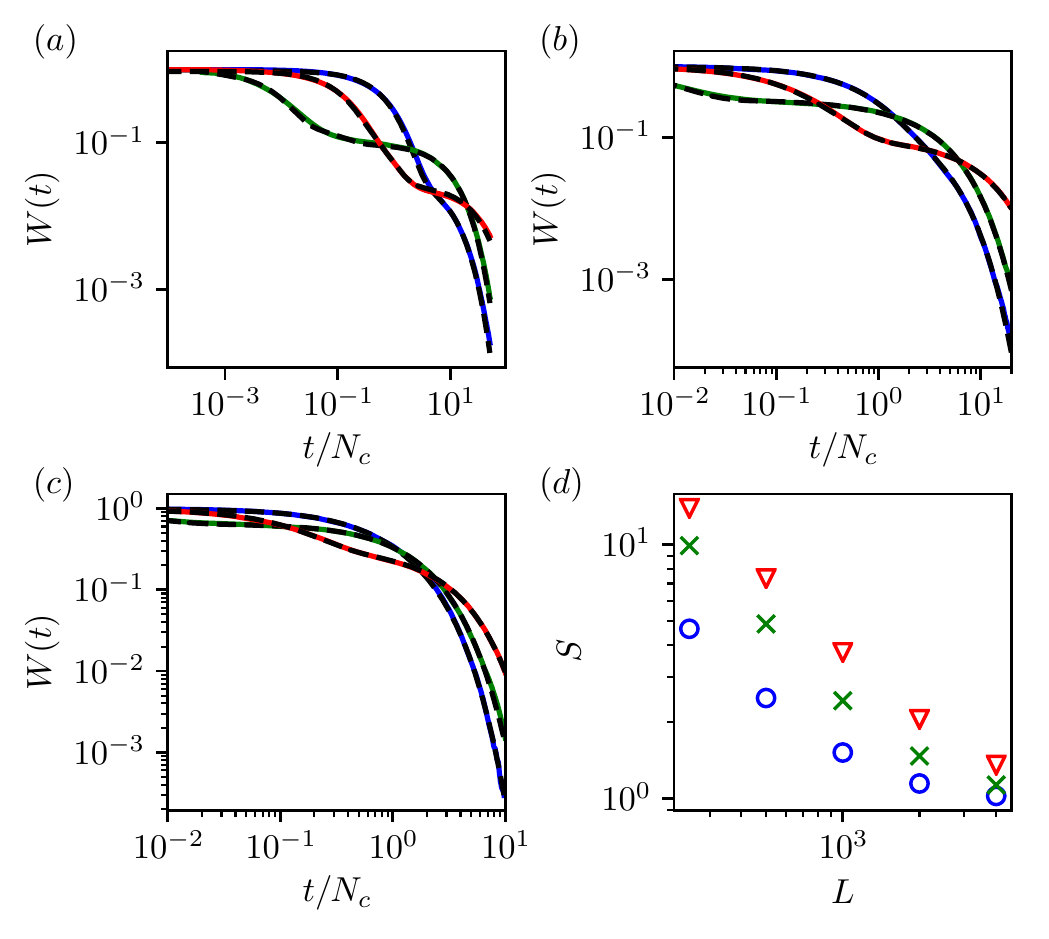}
\par\end{centering}
\caption{\label{fig:Svsgrid} Survival functions for three cells at coordinates 1 (blue): $(0.1,0.1)$, 2 (green): $(0.31,1.94)$, 3 (red): $(0.09,0.2.45)$ for (a) $L=500$, (b) $L=1000$, (c) $L=2000$. The black dashed curves show the fitted hyperexponential distributions \eqref{eq:HyperExponentialDist}. (d) $S$ as a function of grid size in the three cells. Blue circles show cell 1, green crosses cell 2 and red triangles cell 3.}
\end{figure}

Because of the fractal nature of the chaotic sea decreasing the cell size at any scale, resolves more of the underlying structure of the phase space. Typically, new islands will appear with additional sticky sub-structures. In Fig. \ref{fig:Svsgrid} we show the survival functions in the three cells at different grid sizes (a) $L=500$, (b) $L=1000$ and (c) $L=2000$. The parameters of the fitted corresponding hyperexponential distributions are given  in table \ref{table:1} for $L=1000$, table \ref{table:2} for $L=500$ and table \ref{table:3} for $L=500$. The survival functions are not independent of the grid size. While the general shape can be described using the hyperexponential distribution the parameters change when making the cell smaller essentially sampling the distribution in a sub-cell of the larger cell. The time scales represented in the survival functions of the smaller cells are generally of the same order of magnitude as in the larger cell occasionally varying up to a factor of ten (compare the values of $\lambda_i$ each cell at different cell sizes in the tables). The mixing coefficients may, however change significantly when the cell size is changed. Usually one or two of the exponential functions are dominant with the others contributing only a few percent to the mixture. The scaling of $S$ with regard to the grid size is depicted on panel (e). The scaling is not algebraic but is still monotonous and the hierarchy of stickiness is maintained (regions with larger $S$ at some grid size have larger values of $S$ at different grid sizes as well).

In Ref. \citep{meiss1994} Meiss uses the density of orbit visits to show how orbits tend to accumulate around sticky objects for extended periods of time. The paper includes a color plot similar to the $S$-plots. In our experience using the variable $S$ instead of the density gives more stable results both in terms of taking different initial conditions as well as in terms of the number of iterations. The results when using densities can also be somewhat asymmetric  for long times (the densities near the sticky islands at positive $p$ are not the same as those around the equivalent islands at negative $p$) whereas the $S$-plots produce very symmetric results (as can be seen in the animations contained in the supplemental material). The characteristic times in the standard map, including recurrence
times, have recently been analyzed by Harsoula et. al. \citep{harsoula2019},
where they observed similar distributions of recurrence times to small
boxes. They have found exponential distributions in the large chaotic
component before and after the critical value of $k$ and distributions
with long power law tails in the small chaotic components that are
separated below the critical value. The difference in their approach
is that they select multiple initial conditions inside the box, which
may contain also regular initial conditions. In our approach no "contamination" of the results due to the presence of regular trajectories may occur. 

\begin{table}[h]
\begin{center}
\begin{tabularx}{0.8\columnwidth} { | >{\centering\arraybackslash}X | >{\centering\arraybackslash}X | >{\centering\arraybackslash}X | >{\centering\arraybackslash}X | >{\centering\arraybackslash}X | }
 \hline
 \multicolumn{5}{| c|}{Cell 1} \\
 \hline
$\lambda_i$ &0.111 & 1.38 & 32.7 & n/a \\
$p_i$ &0.03 & 0.92 & 0.05 & 0 \\
\hline
 \multicolumn{5}{|c|}{Cell 2} \\
 \hline
$\lambda_i$ & 0.100 & 11.9 & 95.8 & 544 \\
$p_i$ &0.10& 0.09 & 0.59 & 0.22 \\
\hline
 \multicolumn{5}{|c|}{Cell 3} \\
 \hline
$\lambda_i$ &0.0355 & 1.94 &8.12 & 162 \\
$p_i$ &0.03 & 0.18 & 0.71 & 0.07 \\
\hline

\end{tabularx}
\caption{Table of parameter values for the hyperexponetial distributions fitted to the survival functions for $L=500$ presented in Fig. \ref{fig:Svsgrid} (a).}
\label{table:2}
\end{center}
\end{table}

\begin{table}[h]
\begin{center}
\begin{tabularx}{0.8\columnwidth} { | >{\centering\arraybackslash}X | >{\centering\arraybackslash}X | >{\centering\arraybackslash}X | >{\centering\arraybackslash}X | >{\centering\arraybackslash}X | }
 \hline
 \multicolumn{5}{| c|}{Cell 1} \\
 \hline
$\lambda_i$ &0.791 & 2.23 & 53.1 & n/a \\
$p_i$ &0.67 & 0.32 & 0.01 & 0 \\
\hline
 \multicolumn{5}{|c|}{Cell 2} \\
 \hline
$\lambda_i$ & 0.609 & 1.2 & 162 & 1480 \\
$p_i$ &0.56& 0.09 & 0.30 & 0.05 \\
\hline
 \multicolumn{5}{|c|}{Cell 3} \\
 \hline
$\lambda_i$ &0.349 & 6.97 &89.2 & n/a \\
$p_i$ &0.32 & 0.64 & 0.35 & 0 \\
\hline

\end{tabularx}
\caption{Table of parameter values for the hyperexponetial distributions fitted to the survival functions for $L=2000$ presented in Fig. \ref{fig:Svsgrid} (c).}
\label{table:3}
\end{center}
\end{table}

The supplemental material \citep{Supp} of this paper includes an
animation of the $S$ plots as the value of the nonlinearity parameter
changes from $k=0.5$ to $k=11$ in steps of $dk=0.0025$. For each
parameter value $S$ is computed for each cell in the $1000\times1000$
grid, with $T=10^{10}$ and the results are compiled into an animation.
In the animation we can observe how the various invariant curves are
destroyed, leaving behind cantori producing sticky areas around 
islands of stability. The stickiness is most pronounced (the value of $S$ is largest)
right after the destruction of the invariant curve as the holes in
the cantorus are very small and consequently the flux through it is small as well. 
When the chaotic orbit penetrates the area bounded by the cantorus it is trapped between it and the next
invariant curve that survived the perturbation. As the nonlinearity
parameter is increased the holes in the cantorus grow larger, the flux increases, until
eventually the cantorus loses its ability to impede transport and
the sticky area disappears. In the beginning of the animation ($k<k_{c}$)
we see the successive breaking of the invariant tori. In fact it is
still very hard to pinpoint the exact point at which the torus breaks
as the orbit may need a very large number of iterations to find its
way through one of the holes of the cantorus. If the number of numerical
iterations is shorter we may sometimes miss the exact point of the
breaking of the torus. The occasional flickering (a portion of the chaotic component disappears for a few frames of the animation) of the outer most
areas (in the $p$ direction) of the chaotic component is a consequence
of this effect. We also see an overall increase of $S$ even in the
middle of the chaotic sea at the points where chaotic components of
significant sizes merge with the large chaotic sea, either when a
spanning invariant curve, like the golden invariant circle or one of the large islands  breaks apart.
This is very noticeable at small values of $k$ because the overall
relative size of the chaotic sea is small and comparable with the
sizes of the other chaotic components before the merger. The times
for the transitions between the different weakly coupled areas may
be very large. At $k=0.9725$ the time an orbit needs for the transition into the upper part of the phase space may be larger than $10^{10}$ iterations. In Ref. \citep{mackay1984b} the authors give formula for the average transit time trough the golden cantorus $T \sim 25(k-k_c)^{-3.01}$, which gives an estimate of $T = 3.7 \times10^{10}$ at  $k=0.9725$.  The same scaling law holds for any so called boundary circle. 

\section{Results for billiard systems\label{sec:Billiards}}

The other example we provide in this paper are two families of dynamical
billiards. A billiard is a dynamical system which consists of a free
moving particle confined inside a closed domain $\mathcal{B}$ in Euclidean
space referred to as the billiard table. The billiard tables presented
in this paper will all be two dimensional $\mathcal{B}\subset\mathbb{R^{\textrm{2}}}$.
The particle moves freely inside the billiard table in straight lines
and is specularly reflected when hitting the edge of the table, meaning
the angle of reflection is equal to the angle of incidence. The dynamics
can be described as a mapping $\phi$ that gives the position on the
boundary and the velocity of the particle at each successive collision.
As the energy of the particle is conserved the speed can be fixed
to $v=1$ without loss of generality. The phase space can be described
by the Poincar\'e-Birkhoff coordinates $\left(s,\,p\right)$, where
$s$ is the arc-length of the billiard boundary and the conjugated
momentum is the sine of the reflection angle of the particle $p=\sin\alpha$.
The phase space is thus a cylinder $\left(s,\,p\right)\in\left[0,\,\mathcal{L}\right]\times(-1,\,1)$,
where we take $s$ to be periodic with a period equal to the total
length of the billiard boundary $\mathcal{L}$. The dynamics is given
by a sequence of points generated by the area preserving mapping $\left(s,\,p\right)\rightarrow\phi\left(s,\,p\right)$
that maps one collision to the next \citep{Ber1981}.

\subsection{The Robnik billiards}

\begin{figure}[h]
\begin{centering}
\includegraphics[width=1\columnwidth]{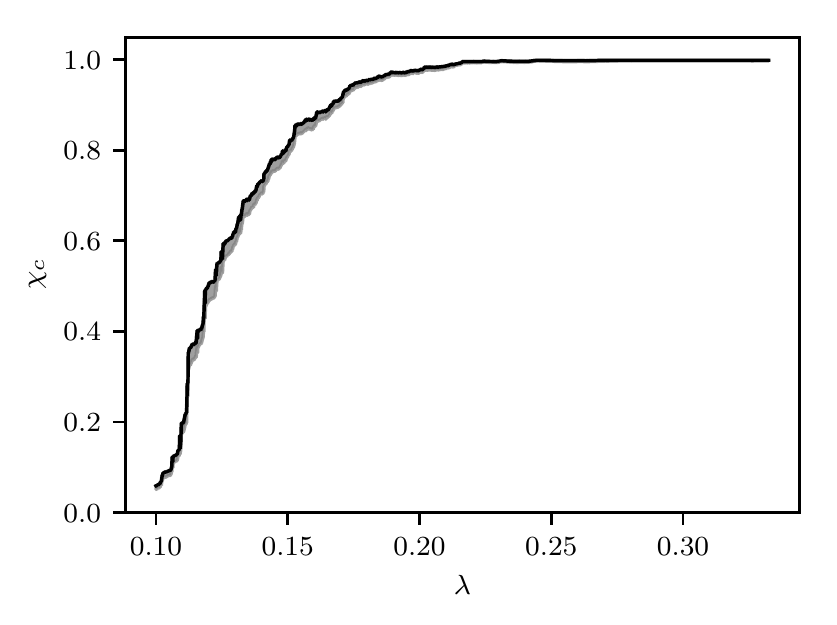}
\par\end{centering}
\caption{\label{fig:RobnikArea}The relative size of the largest chaotic component
$\chi_{c}$ in the Robnik billiards as a function of $\lambda$. The
grid size is $L=1000$ and the orbit was iterated $T=10^{10}$ times.
The gray area shows the error estimated from the number of cells bordering
regular components. See the animation in the supplemental material
for the corresponding $S$-plots.}
\end{figure}

We first present the results for the family of billiards introduced
by Robnik in Ref. \citep{robnik1983} given as a smooth conformal
mapping $z\rightarrow z+\lambda z^{2}$ of the unit disk $|z|=1$
in the complex plane. In the real plane the boundary of the billiards
may be given as the following curve in polar coordinates
\begin{equation}
\boldsymbol{r}\left(\varphi\right)=1+2\lambda\cos\left(\varphi\right),
\end{equation}
 where $\varphi\in\left[0,\,2\pi\right]$ is the polar angle and $\lambda\in\left[0,\,0.5\right]$
is the deformation parameter. We choose the point $\varphi=0$ as
the origin for the arc-length coordinate $s$. The family of billiards
has been well studied both in the classical and quantum domain \citep{robnik1983,robnik1984,BatRob2013B,lozej2018}.
At $\lambda=0$ the boundary is a circle giving an integrable billiard.
The other extreme case $\lambda=0.5$ is the cardioid billiard which
was proven to be an ergodic K-system by Markarian \citep{Markarian1993}.
In between the phase space is generally divided into chaotic components
and regular components. Up to $\lambda=0.25$ regular spanning invariant
curves, known as Lazutkin tori exist because of the convex shape and
the smoothness of the billiard \citep{Laz1973}. Small islands
may remain up until $\lambda=0.5$ but take up only a tiny fraction
of the phase space. In Fig. \ref{fig:RobnikArea} we show how the
relative size of the largest chaotic component changes with the value
of $\lambda$. The curve is qualitatively similar to the one for the
standard map but with less pronounced oscillations. The error has been estimated in the same way as with the standard map described in the previous section.

The supplemental material \citep{Supp} of this paper includes an
animation of the $S$-plots of the largest chaotic component starting
at $\lambda=0.1$ up to $\lambda=0.33$ in steps of $d\lambda=0.0025$
on a grid of $1000\times1000$ cells taking $T=10^{10}$ iterations.
In the animation we see the successive breaking of the invariant tori
that separate the largest chaotic component (forming around the unstable
period-2 orbit) and the smaller chaotic components around the higher
order unstable periodic orbits, following the typical KAM like scenario.
Again, as in the standard map example, we see multiple cantori affecting
the transport in the $p$ direction with $S$ increasing in several
layers towards the edge of the chaotic component. We may also find
cases where stickiness around the islands is dominant. 

\begin{figure}[h]
\begin{centering}
\includegraphics[width=1\columnwidth]{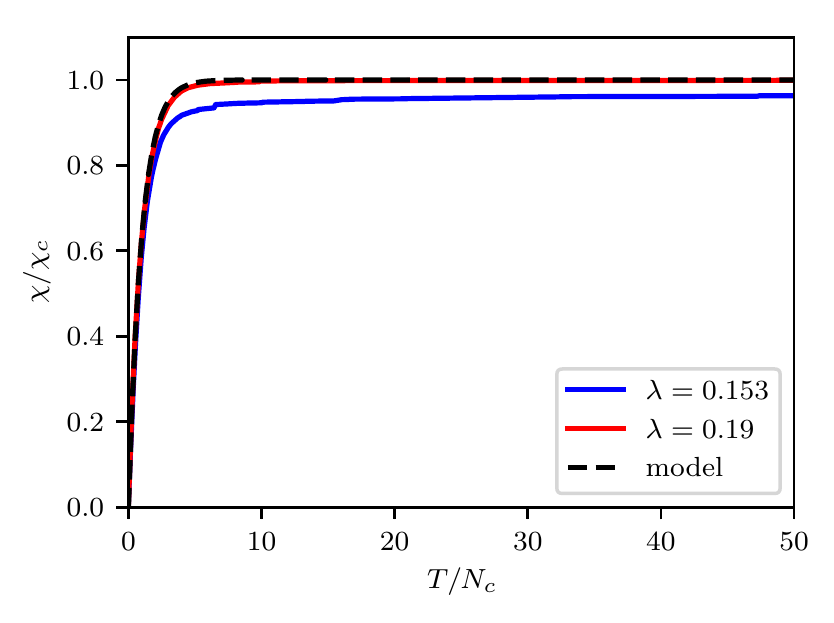}
\par\end{centering}
\caption{\label{fig:CellFillRobnik}The proportion of filled cells (normalized
by the proportion of chaotic cells) with the number Robnik billiard
map iterations (normalized by the number of chaotic cells). The colored
lines show the cell filling curves for different values of the parameter
$\lambda$. Each curve shows the cell filling for a a single chaotic
orbit. The dashed black curve shows the random model prediction. $L=1000$.}
\end{figure}
 
\begin{figure*}[]
\begin{centering}
\includegraphics[width=1\textwidth]{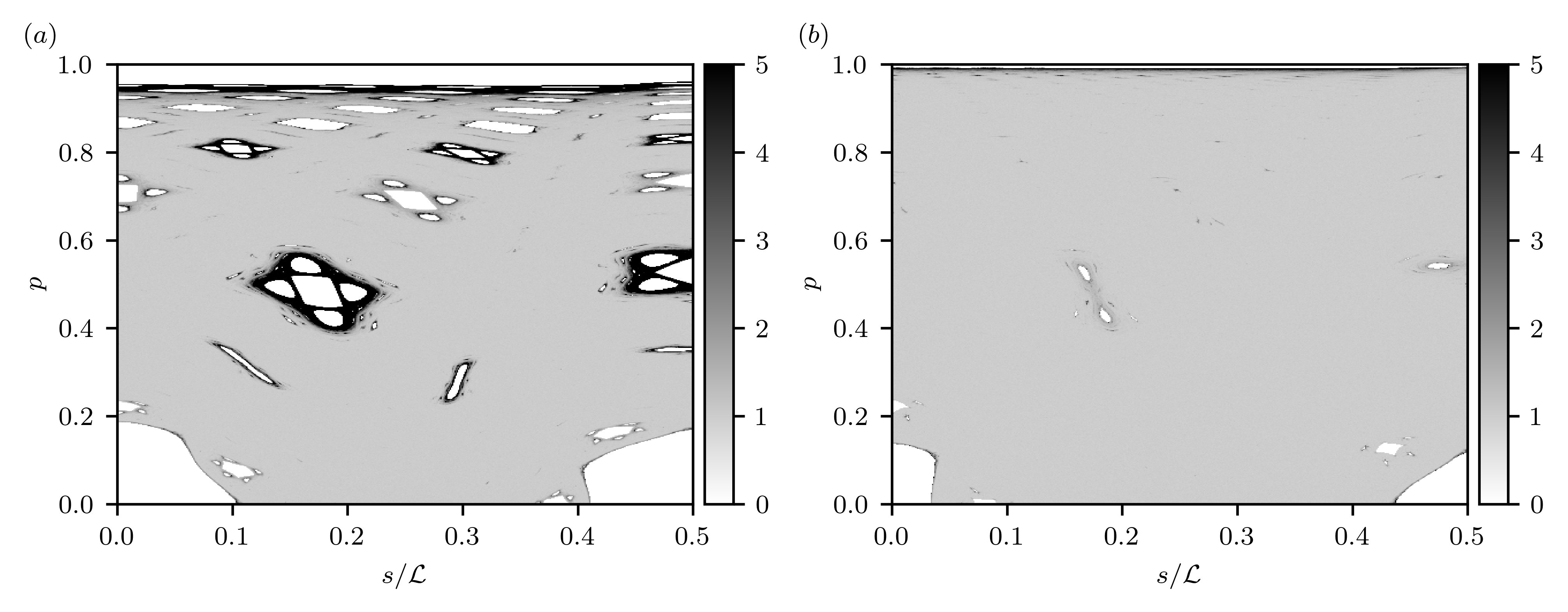}
\par\end{centering}
\caption{\label{fig:RobnikPortraits}The $S$-plots of the largest chaotic
component in the Robnik billiards at (a) $\lambda=0.153$ and (b)
$\lambda=0.19$. Due to symmetry only a quarter of the phase space
is shown $\left(s,\,p\right)\in\left[0,\,\mathcal{L}/2\right]\times\left[0,\,1\right]$. 
The color bar shows the corresponding value of $S$.
Darker areas indicate stickiness. White areas belong to separate invariant
components.}
\end{figure*}

In Fig. \ref{fig:CellFillRobnik} the cell filling graphs for two
$\lambda$ are shown together with the random model prediction. At
$\lambda=0.19$ the numerical curve coincides with the model while
in the $\lambda=0.153$ case the numerical cell filling is slower.
The steps in the curve are very shallow indicating the trapping areas
are relatively small compared to the whole extent of the chaotic component.
The trapping times are also very long. The final extent of the chaotic
component is filled only after $T\approx400\,N_{c}$. The $S$-plots
corresponding to the two parameter values are portrayed in Fig. \ref{fig:RobnikPortraits}.
In panel (a) at $\lambda=0.153$ we see strongly sticky areas around
several KAM islands as well as near the Lazutkin tori. In panel (b)
at $\lambda=0.19$ we still see some regular islands of significant
size but there are no noticeable strongly sticky areas (there might
be a very thin slightly sticky area around the largest regular islands).
The recurrence time distributions are exponential virtually everywhere
in the chaotic component and the motion is uncorrelated. The basic assumption
in the random model is thus largely satisfied which explains agreement
of the numerical cell filling with the model.

\begin{figure}[h]
\begin{centering}
\includegraphics[width=1\columnwidth]{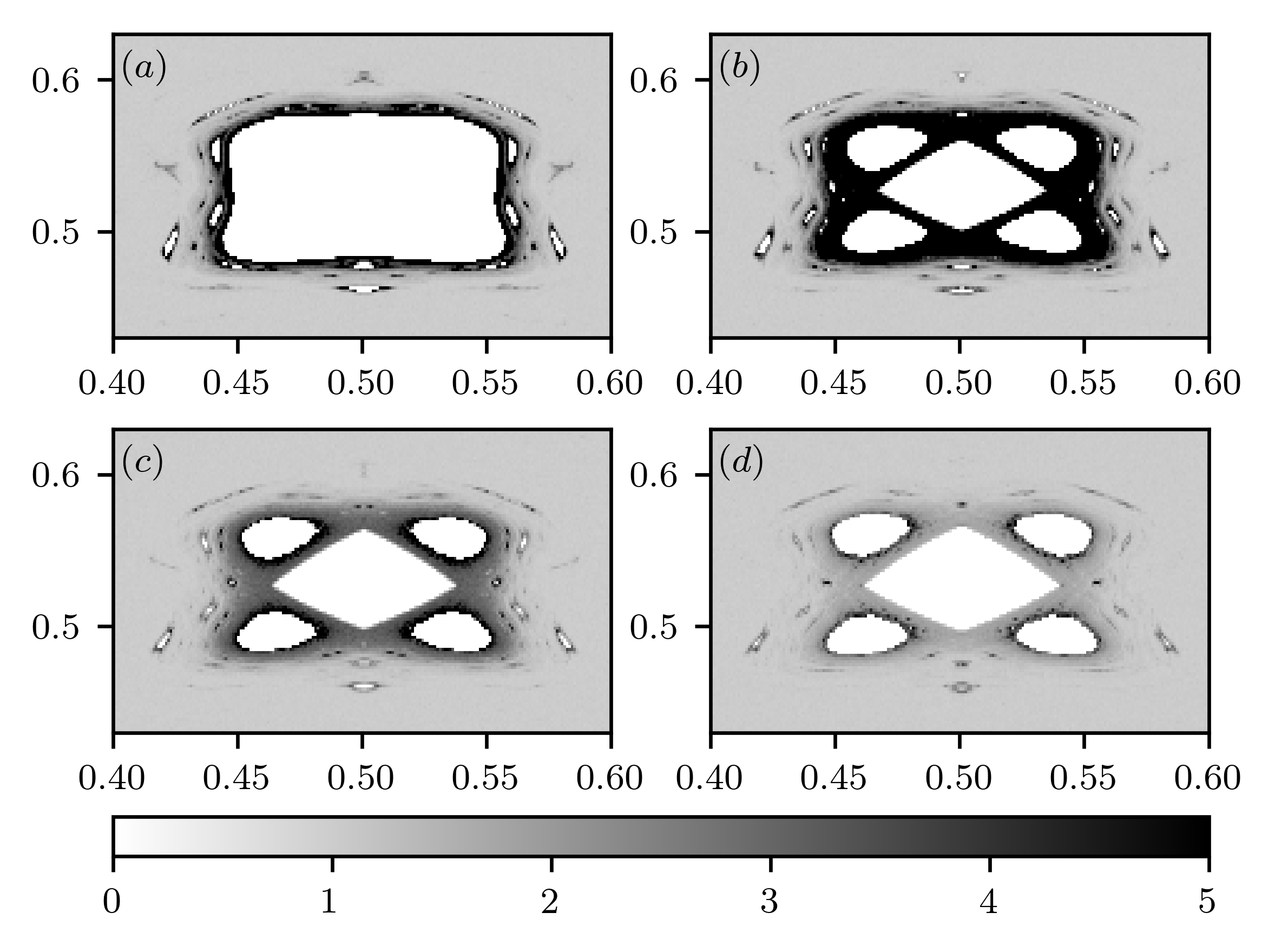}\caption{\label{fig:StickyIslandRobnik}The $S$-plot of the chaotic component
in the vicinity of the islands of stability around the stable period-3 orbit
in the Robnik billiard at (a) $\lambda=0.152$, (b) $\lambda=0.153$, (c) $\lambda=0.154$,
(d) $\lambda=0.155$. The color bar indicates the corresponding values of $S$.}
\par\end{centering}
\end{figure}

In Fig. \ref{fig:StickyIslandRobnik} we show an example of an island,
which surrounds the stable period-3 orbit, breaking because of the
perturbation. We show the $S$-plot of the same area $(s/\mathcal{L},p) \in [0.4,0.6]\times[0.43,0628]$ 
(half of the island can be seen also in Fig. \ref{fig:RobnikPortraits}(a))
at increasing values of the parameter. In panel (a) $\lambda=0.152$
the island is still intact and surrounded by a thin sticky layer and
the typical structure of ever smaller islands around islands of higher
order resonances. In panel (b) $\lambda=0.153$ the island has broken
into a five island structure. The area between the islands is filled
by the chaotic component and is very sticky because of the cantorus
left behind the recently destroyed invariant curve. In panel (c) $\lambda=0.154$
the area between the islands is already far less sticky as the holes
in the cantorus have grown larger. We see thin layers of greater stickiness
around the four flanking islands. In panel (d) $\lambda=0.155$ the
stickiness is almost gone, with only traces remaining. In all four
images we can observe tiny islands of stability around the main islands.
Some of the islands are smaller than the resolution of the cell grid.
If the cell is not entirely filled by the chaotic component the probability
of hitting the cell is smaller. This changes the distribution of recurrence
times to the cell and as a consequence $S$. The $S$-plot is thus
also a good way of detecting tiny islands of stability.

\subsection{The lemon billiards}

\begin{figure}[h]
\begin{centering}
\includegraphics[width=1\columnwidth]{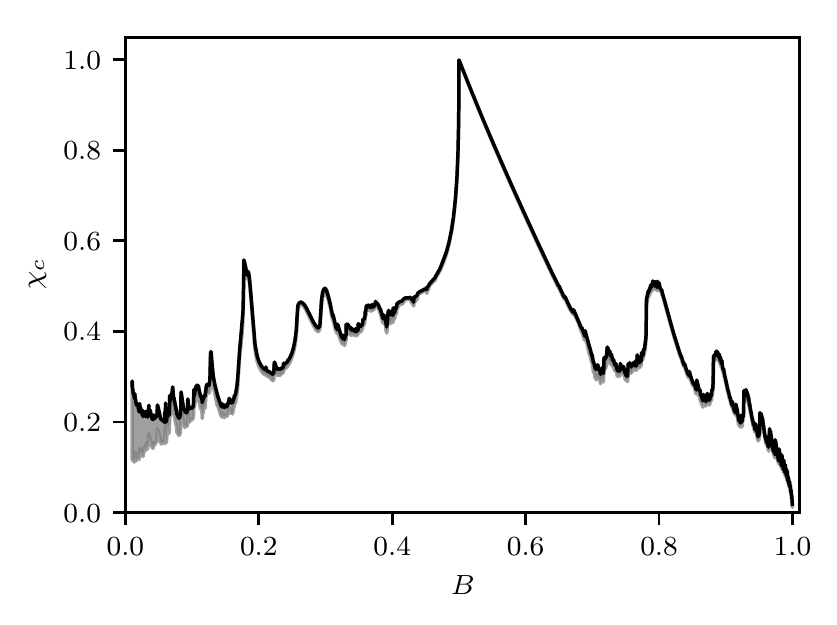}
\par\end{centering}
\centering{}\caption{\label{fig:LemonArea}The relative size of the largest chaotic component
$\chi_{c}$ in the lemon billiards as a function of $B$. The grid
size is $L=1000$ and the orbit was iterated $T=10^{10}$ times. The
gray area shows the error estimated from the number of cells bordering
regular components. See the animation in the supplemental material
for the corresponding $S$-plots.}
\end{figure}

Finally we present the results in the family of billiards known as
the lemon billiards introduced by Heller and Tomsovic in Ref. \citep{HellTom1993}
and further studied together with some generalizations in the classical
and quantum domain by many authors \citep{Lopac1999,MakHarAiz2001,Lopac2001,ChMoZhZh2013,BunZhZh2015}.
The lemon billiard tables are formed by the intersection of two circles
of equal radius (we set $R=1$ without loss of generality) with the
distance between their centers $2B$ being less than their diameters
$B\in(0,1)$. The billiard boundary in the real plane may be given
by the following implicit equations 
\begin{align}
\left(x+B\right)^{2}+y^{2}=1, & x>0,\\
\left(x-B\right)^{2}+y^{2}=1, & x<0.\nonumber 
\end{align}
We choose the point $\left(x,\,y\right)=\left(0,\,-\sqrt{1-B^{2}}\right)$
as the origin for the arc-length coordinate $s$. In contrast to the
Robnik billiards the boundary of the lemon billiards is never smooth
as a kink is formed where the two circular arcs meet. Because of this
there are no Lazutkin tori. The period-2 orbit connecting the points
at the middle of the two circular arcs, $(1-B,\,0)$ and $\left(-1+B,\,0\right)$,
is stable for all values of $B$ with the exception of $B=0.5$ where
it is only marginally stable. In this case this orbit is part of a
whole one-dimensional family of marginally unstable periodic orbits
(MUPO). It is easy to see that at $B=0.5$ any orbit starting from
the middle of the circle will hit the other circle perpendicularly
and retrace its path, because the centers of one circle exactly overlaps
the arc of the other. The phase space of the lemon billiards is thus
of the mixed-type for all values of $B$ with the possible exception
of $B=0.5$, where it might be ergodic. Our numerical results were
not able to verify this as a tiny island, that the chaotic orbit was
unable to penetrate, remained even for $10^{11}$ iterations. In Fig.
\ref{fig:LemonArea} we show how the area of the largest chaotic component
$\chi_{c}$ changes with the value of $B$. The curve bears no resemblance
with the previous examples exhibiting extremely non monotonic behavior.
Because of the small overall size and the large proportion of cells
bordering regular components the error of the estimate is very large
for $B<0.1$.

\begin{figure}[h]
\begin{centering}
\includegraphics[width=1\columnwidth]{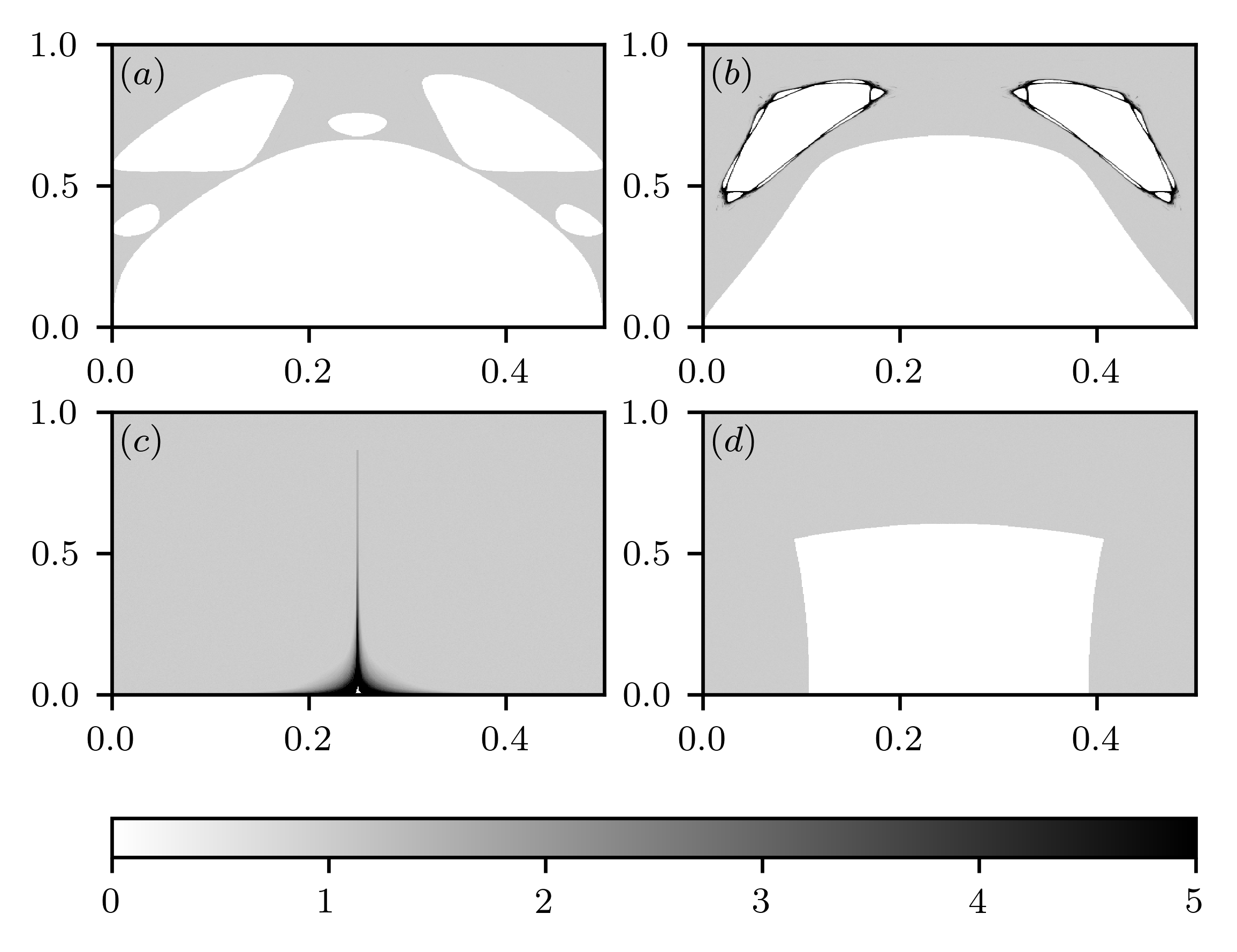}
\par\end{centering}
\caption{\label{fig:LemonPortraits}The $S$-plots of the largest chaotic component
in the lemon billiards at (a) $B=0.24025$, (b) $B=0.31875$, (c)
$B=0.5$ and (d) $B=0.6$. Due to symmetry only a quarter of the phase
space is shown $\left(s,\,p\right)\in\left[0,\,\mathcal{L}/2\right]\times\left[0,\,1\right]$. 
The color bar shows the corresponding value of $S$.
Darker areas indicate stickiness. White areas belong to separate invariant
components.}
\end{figure}

\begin{figure}[h]
\begin{centering}
\includegraphics[width=1\columnwidth]{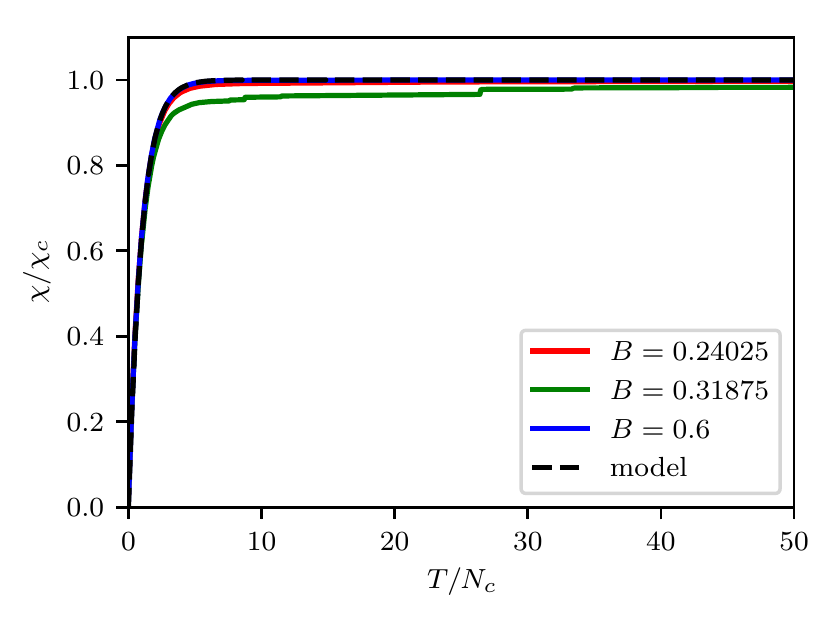}
\par\end{centering}
\caption{\label{fig:CellFillLemon}The proportion of filled cells (normalized
by the proportion of chaotic cells) with the number lemon billiard
map iterations (normalized by the number of chaotic cells). The colored
lines show the cell filling curves for different values of the parameter
$B$. Each curve shows the cell filling for a a single chaotic orbit.
The dashed black curve shows the random model prediction. $L = 1000$.}
\end{figure}

An animation of the $S$-plots of the largest chaotic component for
the lemon billiards is included in the supplemental material \citep{Supp}.
The parameter changes from $B=0.01$ to $B=0.99975$ in steps of $dB=0.00025$.
The grid size is $1000\times1000$ cells and we take $T=10^{10}$
iterations. At $B=0.01$ the billiard shape is very close to a circle.
Only initial conditions that hit the boundary close to the kink where
the two circles meet generate chaotic motion. This results in a very
regular web like structure of the chaotic component and the embedded
KAM islands. The structure can be related to initial conditions that
hit the kink after 1, 2, 3... iterations. When $B$ is increased,
more and more of the periodic orbits lose stability (although there
is a tendency for them to re-stabilize) sometimes leaving behind interesting
sticky structures. The structure of the chaotic components changes
radically with the value of $B$ and many interesting special
cases may be found - we present a selection of them in Fig. \ref{fig:LemonPortraits}.
In panel (a) we see the $S$-plot at $B=0.24025$. Only a few islands
of stability may be seen none of which exhibit any stickiness, with
$S=1$ in all cells belonging to the chaotic sea. At $B=0.31875$,
panel (b), the major islands of stability around the stable period-2
orbit are not sticky, while the the island structures around the stable
period-4 orbit are. In panel (c) we show the special case $B=0.5$
where the sticky family of MUPOs can be observed. The value of $S$
increases roughly exponentially in the sticky area as we get closer
to the MUPO in the star like structure. An exponential increase of
stickiness as one gets closer to the sticky object has been described
by Contopoulos and Harsoula in terms of escape times in Ref. \citep{contopoulos2010a}.
Similar star like structures have been observed by Chen et. al. \citep{ChMoZhZh2013}
in ergodic generalizations of the lemon billiards. A tiny island is
still visible in the middle but we expect this to also be filled if
the orbit is iterated for long enough. For $B>0.5$ the period-2 orbit
is again stable and an island is formed around it. In panel (d) we
show the $S$-plot at $B=0.6$. Only one island of stability is visible
and it is not sticky. Many other interesting examples may be found
in the animation in the supplemental material. The examples with finitely
many islands of stability may prove interesting for more rigorous
analytical treatment. In Fig. \ref{fig:CellFillLemon} the corresponding
cell filling graphs are shown. We see that the cell filling is influenced
by the sticky islands $B=0.31875$. In the other three cases the cell
filling is close to the random model prediction. We see that non-sticky
islands even of substantial sizes like in the cases of $B=0.6$ and
$B=0.24025$ do not induce correlation in the cell visits. The MUPOs
in the $B=0.5$ case also have an almost negligible effect even though
we see stickiness in the $S$-plot. This might be because the sticky
area is very small compared to the size of the chaotic sea. 

\begin{figure}[h]
\begin{centering}
\includegraphics[width=1\columnwidth]{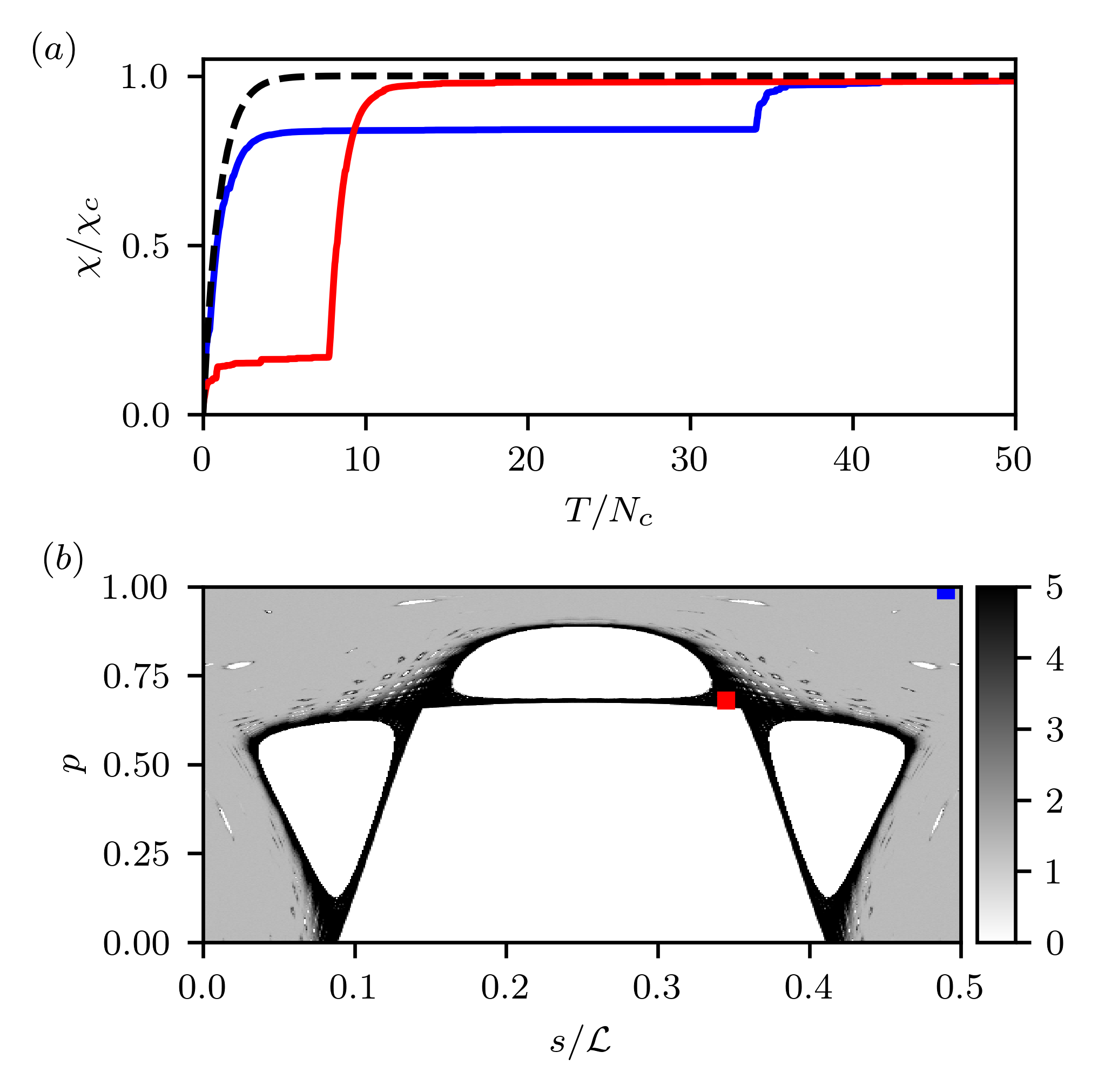}
\par\end{centering}
\caption{\label{fig:LemonSticky}The cell filling in the lemon billiard at
$B=0.78125$. (a) The proportion of filled cells (normalized by the
proportion of chaotic cells) with the number of iterations for two
chaotic orbits, one starting inside the sticky island (red) and the
other outside (blue). (b) The $S$-plot corresponding to the blue
orbit (the other orbit produces practically the same result). The
initial conditions of the two orbits are shown by two boxes of the
same color as the cell filling curve. $L = 1000$.}
\end{figure}

One of the largest sticky areas in this billiard family may be found
at $B=0.78125$. The size of sticky area is approximately $0.2\chi_{c}$.
In Fig. \ref{fig:LemonSticky} we show (panel (a)) the cell filling
for two different orbits, the first started with an initial condition
outside the sticky area and the second inside the sticky area, as
well as the corresponding $S$-plot (panel (b)). The $S$-plot is
the same for both initial conditions. Outside the sticky area $S=1.8$
which is above the expected $S=1$ for the exponential distribution
of recurrence times. This is similar to the result in the standard
map slightly above the critical value of the nonlinearity parameter.
Inside the sticky area the value of $S$ quickly increases and then
plateaus at about $S=12$. The increase is roughly exponential but
is hard to determine exactly as the border of the sticky area is riddled
by islands of stability. The orbit starting from outside the sticky
area (blue) needs a little less than $T=35N_{c}$ iterations to penetrate
inside while the orbit starting from the inside (red) needs a little
less than $T=10N_{c}$ to escape. The escape time is thus much shorter
than the entry time to the sticky area. The shape of the steps in
the two cell filling curves are reminiscent of two periods of exponential
filling of the type given by Eq. \eqref{eq:CellFill}. With some simplification
the system could be described with a two component random model in
the manner of Ref. \citep{RobProDob1999}.

\section{Conclusions and discussion\label{sec:Discussion}}

In this paper we presented a method for analyzing stickiness in chaotic
components of Hamiltonian systems with divided phase space. The method
is based on the examination of recurrence times of a long chaotic
orbit into small cells dividing the phase space. The variable, $S$
which is the ratio between the standard deviation and the mean of
recurrence times, is used to assess the distributions of recurrence
times in the chaotic component. Where $S=1$ the distribution is exponential
and the recurrences are effectively random. In sticky areas a separation of time scales between recurrences occurs due to the dynamical trapping and $S>1$. 

We applied the method to three example systems: the standard map,
the Robnik billiards and the lemon billiards. The main conclusions
are as follows: The random model of diffusion in chaotic components
describes the filling of the cells well in systems with divided phase
space, even when the regular components are of significant size, if
there is no stickiness. Where sticky objects are present the cell
filling is slowed by the cantori causing the stickiness. In the vast
majority of cases $S=1$ in the bulk of the chaotic sea, meaning the
recurrences are completely uncorrelated. The distributions of recurrence
times in these areas are exponential. $S$ rapidly increases in areas
of stickiness in the vicinity of sticky objects. These can be zero
measure objects like sticky marginally unstable periodic orbits
or more extensive object like sticky islands.
The distributions of recurrence times in sticky areas may effectively be modeled using the hyperexponential distribution. When particularly strong
cantori are present in the system that separate large areas of the
chaotic component the value of $S$ is slightly increased even in
the bulk of the chaotic sea, for instance right after the destruction
of a spanning invariant torus. The two examples given in the paper
where this is most visible is the standard map at $k=1$ and the lemon
billiard at $B=0.78125$. The latter example also shows that the time
to escape a sticky region is shorter than the time needed to enter
the sticky region. The $S$-plots provide an excellent overview and
allow us to follow the changes in the structure of the chaotic sea
as we change the parameter. In this way we may identify the positions
of sticky objects as well as the extent and relative strength of the
stickiness. The shape of the sticky areas bordered by cantori can
be seen very clearly. Furthermore, the even tiny islands of stability
that are smaller than the cells can be traced, as they still affect
the recurrence time distributions. The method may therefore also be
useful in providing numerical evidence of ergodicity. 
The statistic $S$ is stable with regards to the initial condition of the chaotic orbit and the number of iterations after a transient regime.
However $S$ is not stable with regard to the grid size, which is a significant drawback. The scaling of $S$ is not algebraic but still monotonous and the $S$-plots at different grid sizes give qualitatively identical results. 

As mentioned in the introduction there has been a long standing debate about the presence and universality of algebraic decay of recurrence time distributions in Hamiltonian systems with divided phase space. Our method only distinguishes an exponential distribution from any other distribution. This might be a hyperexponential distribution as was the case in all the examples found in this paper or any other distribution including those with power law tails.  However,  detecting distributions with algebraic decay exponents $\gamma < 3$ is possible with our method, as the standard deviations for such distributions diverges. If the value of $S$ keeps increasing with the number of iterations this would be a very clear indication of such an algebraic decay. It would also be interesting to study the recurrences in specific sticky areas as a unified domain. The $S$-plots may be used to determine the borders of the sticky area of interest and then the escape times or recurrence times to the specified domain can be studied. One of the interesting open questions is also how to quantitatively assess the effects of stickiness on the transport inside the chaotic component, especially in cases where the phase space is bounded like in billiard systems where the usual arguments using the decay of correlations \citep{karney1983} have to be modified \citep{venegeroles2009}. The sticky areas may slow down transport at finite times considerably, leading to anomalous diffusion \citep{manos2014}. As recently shown for the standard map \citep{harsoula2018} the transition to the asymptotic regime may be extremely long.    

We used the $S$-plots to identify several interesting cases in the
lemon billiard with apparently finitely many islands of stability and no stickiness
as well as cases with extreme stickiness. Such special cases might
lend themselves to more rigorous mathematical analysis paving the
way to a more general understanding of the transport in mixed type
systems in the generic case, which is a long standing open problem.
The special cases in the lemon billiards may also prove interesting
for analysis in the quantum domain as the spectral statistics and
localization of eigenstates in the semiclassical limit are closely
linked to the transport properties of the classical system \citep{BatLozRob2019}. 
Research of the effects of stickiness on the localization of quantum eigenstates
is currently underway. 

\section*{Acknowledgments}

The author acknowledges the financial support from the Slovenian Research
Agency (research core funding No. P1-0306). The author would like
to thank Prof. Marko Robnik for support and stimulating discussions, and careful reading of the manuscript
and Dr. B. Batisti\'c for providing the use of his excellent numerical
library \citep{Benokit}. The author thanks the anonymous referees for excellent critical, informative and helpful remarks.

\providecommand{\noopsort}[1]{}\providecommand{\singleletter}[1]{#1}

\end{document}